\pdfoutput=1
\documentclass[10pt,a4paper,oneside,linenumbers=off,latinmodern=off,timesnews=off,english]{rctart-class/rctart}

\setboolean{article-info}{false}              % false to hide the article-info section
\setboolean{tarja-info}{false}                % false to hide the article-info section
\setboolean{articleinfotop}{false}            % false to print article-info after affiliations
\setboolean{colorabst}{false}                 % colore o resumo e abstract
\setboolean{printhead}{false}
\setboolean{printfoot}{true}

\journalname{Semana da Facet - 2024}
\title{Perturbation-based Nonperturbative Method}

\author[1]{Chang Liu}
\author[2]{Wen-Du Li}
\author[1]{Wu-Sheng Dai}

\affil[1]{Department of Physics, Tianjin University, Tianjin 300350, P.R. China}
\affil[2]{College of Physics and Materials Science, Tianjin Normal University, Tianjin 300387, PR China}

\dates{}%Este manuscrito foi compilado em 28 de abril de 2024}

\footinfo{Annals of Physics 468 (2024) 169741}%{Semana da Facet -- 2024}
\smalltitle{Modelo \LaTeX\ para a \href{https://periodicos.unemat.br/index.php/recet}{Recet}}
\institution{Universidade do Estado de Mato Grosso}
\theyear{2024}
\thevolume{1}
\thelocal{Sinop}
\themonths{Jul.-Dez.}
\theyears{2023/Jan.-Dez.2024}
\corres{Forneça as informações do autor e editor correspondentes aqui.}
\email{example@organization.com.}
\doi{\url{https://www.doi.org/exampledoi/XXXXXXXXXX}}
\received{YY/ZZ/2024}
\revised{YY/ZZ/2024}
\accepted{YY/ZZ/2024}
\published{YY/ZZ/2024}
\eissn{2965-9558}
\articlenum{e0124XX}

\license{}%Classe Rctart \LaTeX\ \ccLogo\ Este documento está licenciado sob Creative Commons CC BY 4.0.}

\begin{document}
\twocolumn[
    \maketitle
    \thispagestyle{firststyle}
    \linenumbers 
%     \tableofcontents

%----------------------------------------------------------

%----------------------------------------------------------
% ABSTRACT
%----------------------------------------------------------

% \palavraschave{palavra-chave 1, palavra-chave 2, palavra-chave 3, palavra-chave 4, palavra-chave 5}
% \keywords{keyword 1, keyword 2, keyword 3, keyword 4, keyword 5}
]

%\begin{otherlanguage}{english}
%\begin{abstract}
\begin{rctartenv}[frametitle=ABSTRACT]

  This paper presents a nonperturbative method for solving eigenproblems. This
method applies to almost all potentials and provides nonperturbative
approximations for any energy level. The method converts an eigenproblem into
a perturbation problem, obtains perturbation solutions through standard
perturbation theory, and then analytically continues the perturbative solution
into a nonperturbative solution. Concretely, we follow three main steps: (1)
Introduce an auxiliary potential that can be solved exactly and treat the
potential to be solved as a perturbation on this auxiliary system. (2) Use
perturbation theory to obtain an approximate polynomial of the eigenproblem.
(3) Use a rational approximation to analytically continue this approximate
polynomial into the nonperturbative region.
\end{rctartenv}

% \printkeywords{Nonperturbative method $|$ Approximate analytic continuation $|$ Rational approximation $|$ Eigenproblem.} % must put \@keyword or \@wordskey or directly the text of the keywords
%\end{abstract}
%\end{otherlanguage}
% ]
% 
% \begin{otherlanguage}{spanish}
% \begin{abstract}
%   Bienvenido a la clase \cls{rctart} \LaTeX\ para realizar trabajos académicos e informes de laboratorio. En esta plantilla de ejemplo, lo guiaremos a través del proceso de uso y personalización del documento según sus necesidades. Para obtener más información sobre esta clase, consulte la sección de apéndices. Allí encontrarás códigos que definen los aspectos principales del modelo, permitiéndote explorarlos y modificarlos. Si no necesita el resumen, comente el entorno \env{abstract}. Cabe mencionar que esta plantilla está inspirada en el trabajo de otros autores de la clase \LaTeX, la clase \cls{\href{https://memonotess1.wixsite.com/memonotess}{rho}} \LaTeX\, diseñado con intenciones académicas.
%    \printkeywords{palabra clave 1, palabra clave 2, palabra clave 3, palabra clave 4, palabra clave 5.} % debe poner \@keyword o \@wordskey o directamente el texto de las palabras clave
% \end{abstract}
% \end{otherlanguage}

%----------------------------------------------------------

\section{Introduction}

Exact solutions are rare and require approximation methods. In quantum
mechanics, approximation methods can be divided into perturbation methods,
such as the stationary perturbation theory, and nonperturbation methods, such
as the variational method. The advantage of the perturbation method lies in
its standardized procedures, but it is limited to perturbative problems.
Nonperturbation methods lack universally standardized procedures; for
instance, the variational method must choose trial wave functions
artificially, relying entirely on guesswork and lacking a systematic
construction approach. In this paper, we establish a nonperturbation
approximation method to solve the eigenproblem of the Hamiltonian%
\begin{equation}
H=-\frac{\hbar^{2}}{2\mu}\nabla^{2}+V. \label{H}%
\end{equation}
This nonperturbation method has a standardized procedure like that in the
perturbation theory, which provides an explicit nonperturbation expression of
the eigenvalue and eigenfunction.

The basic idea can be briefly divided into two steps:

(1) Converting an eigenproblem, a nonperturbation problem, into a perturbation
problem and using the standard perturbation theory to calculate the
eigenproblem perturbatively.

(2) Analytically continuing the obtained perturbation results into a
nonperturbation result.

The technical key is the analytic continuation. Analytic continuation,
theoretically, can be done as follows: first, calculate a perturbation series
for all orders (which is impractical); then, sum up the perturbation series to
obtain a sum function (which is also impractical). This sum function is just
the analytic continuation of the perturbative series. For instance, if the
perturbation series is a power series, the effective region is a disk with a
radius equal to the convergence radius. The analytic region of the sum
function has an analytic region encompassing a larger area, including that
disk. The circle of convergence of a perturbative power series is where
perturbation theory holds. The analytic region of the sum function outside the
convergence circle is the nonperturbative region. However, in practice, we can
neither obtain every term of a perturbation series nor are we likely to be
able to sum the series.

Perturbation theory usually provides a polynomial approximation, the first
several terms of a perturbation series, rather than the entire series.
Therefore, what we need to do is to analytically continue an approximate
polynomial, rather than a perturbation series, to a nonperturbative result. In
this paper, we will use rational approximation, which approximates a function
by a rational expression, to perform an approximate analytic continuation,
analytically continuing the perturbative polynomial to a nonperturbative region.

The idea of rational approximation was mentioned in Euler's
"\textit{Introduction to Analysis of the Infinite}"
\cite{euler2012introduction}. The rational approximation uses a rational
expression to approximate a function. As a comparison, power series
approximation uses a truncated power series, a polynomial, to approximate a function.

The advantage of power series approximation lies in its operational
convenience, as it has a standardized procedure for calculating the
coefficients in the power series, such as the Taylor expansion in mathematics
and the Feynman diagrammatic expansion and the virial expansion in physics.
The disadvantage of power series approximation is that it is valid only in a
perturbation region.

The advantage of rational approximation is that besides perturbation regions,
it also applies to nonperturbation regions. The disadvantage of rational
approximation is the lack of a standardized procedure for obtaining the
coefficients in the rational function.

The rational approximation adopted in this paper employs Pad\'{e}
approximation \cite{baker1996pade,baker1961pade,baker1975essentials}, where
the coefficients of the rational function approximation are determined by the
polynomial approximation given by perturbation theory. Of course, there are
other methods for rational approximation besides the Pad\'{e} approximation,
such as determining the coefficients of the rational function through
numerical fitting.

Much research has been conducted on solving the Schr\"{o}dinger equation,
including exact solutions, quasi-exact solutions, and approximate methods. For
exact solutions, the supersymmetric quantum-mechanical method
\cite{ahmadov2018analytical,bermudez2013factorization,andrianov2012nonlinear,bermudez2016solutions}%
, the Nikiforov-Uvarov method \cite{kumar2021new,falaye2013exact}, the Laplace
transform method \cite{arda2012exact}, the factorization method
\cite{sanchez2014dirac}, the symplectic quantum mechanical method
\cite{de2018supersymmetric}, and the dual transformation method
\cite{li2021duality,li2022solving} are developed, and the solutions of, e.g.,
$1/\sqrt{r}$-, $r^{2/3}$-, $r^{6}$-, $1/r^{2/3}$-, and $r^{6}$-potential are
obtained \cite{li2016exact,chen2022exact,li2022solving}. In addition to exact
solutions, research has also been conducted on quasi-exact solutions
\cite{agboola2012unified,agboola2013novel,turbiner2016one,hatami2017exact,agboola2014new,xie2011new}%
. For more complex problems, approximate methods have been developed
\cite{amore2005new,dorey2005beyond}.

In section \ref{Method}, we introduce the nonperturbative method. In section
\ref{Eigenvalue}, we calculate the nonperturbative eigenproblem. In section
\ref{auxiliary}, We consider the construction of auxiliary potentials. The
summary is given in section \ref{Conclusion}.

\section{Nonperturbative method}\label{Method}

The method consists of two steps: (1) converting a nonperturbative problem
into a perturbative problem; (2) analytically continuing the obtained
perturbative result to a nonperturbative one.

\subsection{Step 1: Converting nonperturbation to perturbation}\label{step1}

In this step, we convert the eigenproblem of the Hamiltonian (\ref{H}), a
nonperturbation problem, into a perturbation problem.

By introducing an auxiliary potential $U$, we rewrite the Hamiltonian
(\ref{H}) into a perturbative form:
\begin{align}
\mathcal{H}\left(  \lambda\right)   &  =\left(  -\frac{\hbar^{2}}{2\mu}%
\nabla^{2}+U\right)  +\lambda\left(  V-U\right) \nonumber\\
&  =H_{0}+\lambda\Delta, \label{2.19}%
\end{align}
where $H_{0}=-\frac{\hbar^{2}}{2\mu}\nabla^{2}+U$ is exactly solvable. The
eigenproblem of the auxiliary Hamiltonian $\mathcal{H}\left(  \lambda\right)
$,%
\begin{equation}
\mathcal{H}\left(  \lambda\right)  \psi_{n}\left(  \lambda\right)
=\mathcal{E}_{n}\left(  \lambda\right)  \psi_{n}\left(  \lambda\right)  ,
\label{eigenequation}%
\end{equation}
recovers the eigenproblem of the Hamiltonian (\ref{H}) when $\lambda=1$:%
\begin{equation}
H\psi_{n}=E_{n}\psi_{n}%
\end{equation}
with $\mathcal{H}\left(  1\right)  =H$, $\psi_{n}\left(  1\right)  =\psi_{n}$,
and $\mathcal{E}_{n}\left(  1\right)  =E_{n}$.

The eigenproblem of $\mathcal{H}\left(  \lambda\right)  $, when $\lambda<1$,
is a perturbative problem, and, when $\lambda=1$, is a nonperturbative problem.

We first treat $\lambda$ as a small parameter, namely, $\lambda<1$, allowing
us to use perturbation theory to solve the eigenequation (\ref{eigenequation}%
), which gives a polynomial approximation of the eigenvalue and
eigenfunction:
\begin{align}
\mathcal{E}_{n}\left(  \lambda\right)   &  \simeq\mathcal{E}_{n}^{\left(
0\right)  }+\mathcal{E}_{n}^{\left(  1\right)  }\lambda+\mathcal{E}%
_{n}^{\left(  2\right)  }\lambda^{2}+\cdots+\mathcal{E}_{n}^{\left(  N\right)
}\lambda^{N},\label{2.21}\\
\psi_{n}\left(  \lambda\right)   &  \simeq\psi_{n}^{(0)}+\psi_{n}^{(1)}%
\lambda+\psi_{n}^{(2)}\lambda^{2}+\cdots+\psi_{n}^{(N)}\lambda^{N}.
\end{align}
This polynomial approximation, a perturbative result, fails at $\lambda=1$.

\subsection{Step 2: Analytically continuating perturbation to nonperturbation}

In Step 1, perturbation theory yields perturbative polynomial approximations
of the eigenvalue and eigenfunction of the auxiliary Hamiltonian
$\mathcal{H}\left(  \lambda\right)  $. In Step 2, we use rational
approximations to approximately analytically continue the perturbative
polynomial approximation obtained in Step 1 to a nonperturbative one valid at
$\lambda=1$. The nonperturbative approximation of the eigenvalue and
eigenfunction are then $E_{n}=\mathcal{E}_{n}\left(  1\right)  $ and $\psi
_{n}=\psi_{n}\left(  1\right)  $.

The exact analytic continuation is unique, but the approximate analytic
continuation is not, depending on the chosen approximation method. However,
the difference between the approximate analytic continuation obtained from
different rational approximations is often insignificant. For instance, two
methods can be employed to obtain a rational approximation of a polynomial:
one is the Pad\'{e} approximation, and the other fits the polynomial with a
rational expression. If using the Pad\'{e} approximation, the sum of the
number of terms of the numerator and denominator polynomials of the rational
expression must equal that of the polynomial. In contrast, using a rational
expression to fit the polynomial imposes no restrictions on the number of
terms of the rational expression, allowing for rational approximations of
various numbers of terms to approximate the polynomial. Direct attempt shows
that the results yielded by these two methods exhibit only a small discrepancy.

In this paper, we use the Pad\'{e} approximation to perform rational
approximation, which determines the coefficients in the rational approximation
by the perturbative polynomial. The advantage of this method is that both the
perturbation theory in Step 1 and the analytic continuation by the Pad\'{e}
approximation in Step 2 have standard procedures.

\begin{figure}[ptb]
\centering
%Requires \usepackage{graphicx}
\includegraphics[width=0.40\textwidth]{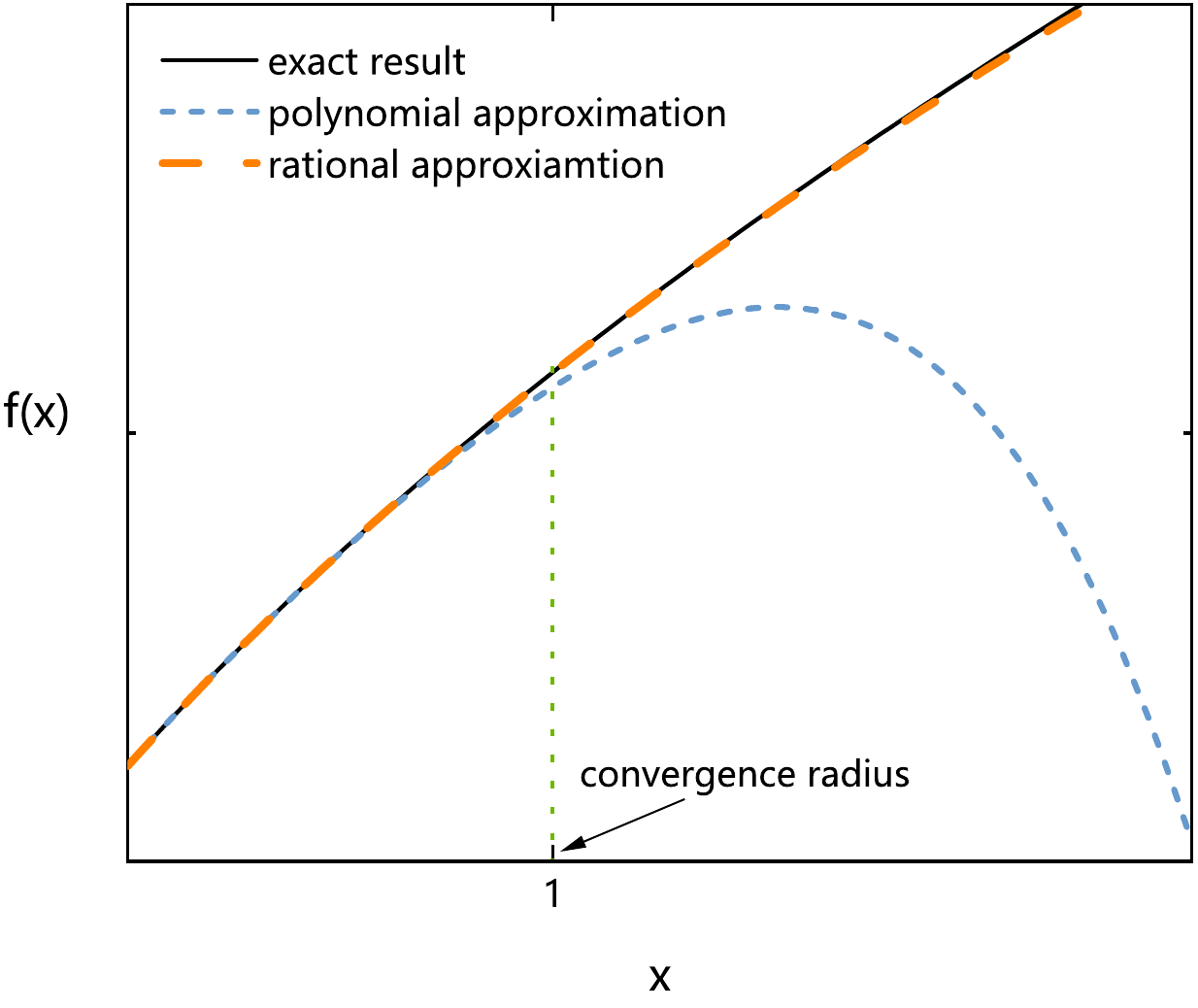}\newline\caption{ Comparison
of polynomial approximation, rational approximation, and exact value of
$f\left(  x\right)  =\sqrt{1+x}$.}%
\label{fx}%
\end{figure}

\textit{Mathematical example}. The convergence radius of the power series
$f\left(  x\right)  =\sqrt{1+x}=\sum_{n=0}^{\infty}\frac{\left(  -1\right)
^{n+1}\left(  2n-3\right)  !!}{2^{n}n!}x^{n}$ is $1$. Assuming that, as in
perturbation theory, we only obtain a quartic polynomial approximation of this
power series:%
\begin{equation}
f^{\text{P}}\left(  x\right)  =1+\frac{x}{2}-\frac{x^{2}}{8}+\frac{x^{3}}%
{16}-\frac{5x^{4}}{128}. \label{1.9}%
\end{equation}
This approximation is only valid for $x<1$ and fails for $x>1$, as shown in
Figure \ref{fx}.

If we approximate the polynomial $f^{\text{P}}\left(  x\right)  $ with a
rational function $f^{\text{R}}\left(  x\right)  $ whose numerator and
denominator are both quadratic polynomials,%
\begin{equation}
f^{\text{R}}\left(  x\right)  =\frac{1+\frac{5}{4}x+\frac{5}{16}x^{2}}%
{1+\frac{3}{4}x+\frac{x^{2}}{16}}, \label{Pex}%
\end{equation}
then for $x<1$, we have $f^{\text{R}}\left(  x\right)  \simeq f^{\text{P}%
}\left(  x\right)  $, so $f^{\text{R}}\left(  x\right)  $ is an approximate
analytic continuation of $f^{\text{P}}\left(  x\right)  $. For $x>1$,
$f^{\text{R}}\left(  x\right)  $ provides a good approximation, as shown in
Figure \ref{fx}.

This example demonstrates how to use a Pad\'{e} approximation to perform an
approximate analytical continuation of a polynomial approximation.

\section{Eigenproblem}\label{Eigenvalue}

\subsection{Perturbative approximation}

Using the standard stationary perturbation theory to solve the eigenequation
(\ref{eigenequation}) gives polynomial approximations of the eigenvalue and
eigenfunction of the auxiliary Hamiltonian $\mathcal{H}\left(  \lambda\right)
$ of the form (\ref{2.21}). The $N$th-order eigenfunction and eigenvalue, by
the stationary perturbation theory, can be calculated directly:%
\begin{align}
\mathcal{E}_{n}^{\left(  N\right)  } &  =\sum_{m\neq n=1}\frac{K\left(
N-1,m,n\right)  }{\mathcal{E}_{n}^{\left(  0\right)  }-\mathcal{E}%
_{m}^{\left(  0\right)  }}\Delta_{mn},\label{recursion1}\\
\psi_{n}^{(N)} &  =\sum_{m\neq n=1}\frac{K(N,m,n)}{\mathcal{E}_{n}%
^{(0)}-\mathcal{E}_{m}^{(0)}}\psi_{m}^{(0)},\text{ \ \ }N\geq2,
\end{align}
where the zero-order eigenfunction and eigenvalue, $\psi_{n}^{\left(
0\right)  }$ and $\mathcal{E}_{n}^{\left(  0\right)  }$, are given by the
exact solution of $H_{0}\psi_{n}^{\left(  0\right)  }=\mathcal{E}_{n}^{\left(
0\right)  }\psi_{n}^{\left(  0\right)  }$ and%
\begin{align}
& K\left(  N,m,n\right)  \label{K-correct}\\
& =\underset{p\neq n}{\sum}\Delta_{mp}\frac{K\left(  N-1,p,n\right)
}{\mathcal{E}_{n}^{\left(  0\right)  }-\mathcal{E}_{p}^{\left(  0\right)  }%
}-\underset{j=1}{\overset{N-1}{\sum}}\mathcal{E}_{n}^{\left(  j\right)  }%
\frac{K\left(  N-j,m,n\right)  }{\mathcal{E}_{n}^{\left(  0\right)
}-\mathcal{E}_{m}^{\left(  0\right)  }}\,,N>1\nonumber
\end{align}
with $\Delta_{mn}=\left\langle \psi_{m}^{\left(  0\right)  }\right\vert
\Delta\left\vert \psi_{n}^{\left(  0\right)  }\right\rangle $ and
$\mathcal{E}_{n}^{(1)}=\Delta_{nn}$.

\subsection{Nonperturbative approximation}\label{solution}

To perform an analytic continuation of the polynomial approximation of the
eigenvalue and eigenfunction obtained by perturbation theory, we approximate
the polynomial (\ref{2.21}) by a rational function:%
\begin{align}
\mathcal{E}_{n}^{\left[  L/M\right]  }\left(  \lambda\right)   &
=\frac{P_{\mathcal{E}}^{[L/M]}(\lambda)}{Q_{\mathcal{E}}^{[L/M]}(\lambda
)}\simeq\sum_{i=0}^{N}\mathcal{E}_{n}^{\left(  i\right)  }\lambda
^{i},\label{PeqR}\\
\psi_{n}^{[L/M]}\left(  \lambda\right)   &  =\frac{P_{\psi}^{[L/M]}(\lambda
)}{Q_{\psi}^{[L/M]}(\lambda)}\simeq\sum_{i=0}^{N}\psi_{n}^{(i)}\lambda^{i}.
\end{align}
When using the Pad\'{e} approximation to perform the rational approximation,
the coefficient of the rational approximation is determined by the polynomial
approximation, in which the numerator
\begin{equation}
P_{a}^{[L/M]}(\lambda)=\left\vert
\begin{array}
[c]{cccc}%
a_{n}^{\left(  L-M+1\right)  } & a_{n}^{\left(  L-M+2\right)  } & \cdots &
a_{n}^{\left(  L+1\right)  }\\
a_{n}^{\left(  L-M+2\right)  } & a_{n}^{\left(  L-M+3\right)  } & \cdots &
a_{n}^{\left(  L+2\right)  }\\
\vdots & \vdots & \ddots & \vdots\\
a_{n}^{\left(  L-1\right)  } & a_{n}^{\left(  L\right)  } & \cdots &
a_{n}^{\left(  L+M-1\right)  }\\
a_{n}^{\left(  L\right)  } & a_{n}^{\left(  L+1\right)  } & \cdots &
a_{n}^{\left(  L+M\right)  }\\
\sum\limits_{i=0}^{L-M}a_{n}^{\left(  i\right)  }\lambda^{M+i} &
\sum\limits_{i=0}^{L-M+1}a_{n}^{\left(  i\right)  }\lambda^{M+i-1} & \cdots &
\sum\limits_{i=0}^{L}a_{n}^{\left(  i\right)  }\lambda^{i}%
\end{array}
\right\vert \label{2.30}%
\end{equation}
is an $L$th-order polynomial and the denominator%
\begin{equation}
Q_{a}^{[L/M]}(\lambda)=\left\vert
\begin{array}
[c]{ccccc}%
a_{n}^{\left(  L-M+1\right)  } & a_{n}^{\left(  L-M+2\right)  } & \cdots &
a_{n}^{\left(  L\right)  } & a_{n}^{\left(  L+1\right)  }\\
a_{n}^{\left(  L-M+2\right)  } & a_{n}^{\left(  L-M+3\right)  } & \cdots &
a_{n}^{\left(  L+1\right)  } & a_{n}^{\left(  L+2\right)  }\\
\vdots & \vdots & \ddots & \vdots & \vdots\\
a_{n}^{\left(  L-1\right)  } & a_{n}^{\left(  L\right)  } & \cdots &
a_{n}^{\left(  L+M-2\right)  } & a_{n}^{\left(  L+M-1\right)  }\\
a_{n}^{\left(  L\right)  } & a_{n}^{\left(  L+1\right)  } & \cdots &
a_{n}^{\left(  L+M-1\right)  } & a_{n}^{\left(  L+M\right)  }\\
\lambda^{M} & \lambda^{M-1} & \cdots & \lambda & 1
\end{array}
\right\vert , \label{2.29}%
\end{equation}
is an $M$th-order polynomial with $N=L+M$ \cite{tian2021pade}, where for the
eigenvalue $a_{n}^{\left(  i\right)  }=\mathcal{E}_{n}^{\left(  i\right)  }$
and for the eigenfunction $a_{n}^{\left(  i\right)  }=\psi_{n}^{(i)}$.

The eigenvalue $\mathcal{E}_{n}^{\left[  L/M\right]  }\left(  \lambda\right)
$ and the eigenfunction $\psi_{n}^{[L/M]}\left(  \lambda\right)  $ obtained in
this way are analytically continued results and remain valid at $\lambda=1$ in
the nonperturbative region. $\mathcal{E}_{n}^{\left[  L/M\right]  }\left(
1\right)  $ and $\psi_{n}^{[L/M]}\left(  1\right)  $ provide nonperturbative
approximative eigenvalues and eigenfunctions for the Hamiltonian (\ref{H}):%
\begin{align}
E_{n}  &  =\mathcal{E}_{n}^{\left[  L/M\right]  }\left(  1\right)
=\frac{P_{\mathcal{E}}^{[L/M]}(1)}{Q_{\mathcal{E}}^{[L/M]}(1)}, \label{EnPade}%
\\
\psi_{n}  &  =\psi_{n}^{[L/M]}\left(  1\right)  =\frac{P_{\psi}^{[L/M]}%
(1)}{Q_{\psi}^{[L/M]}(1)}. \label{psiPade}%
\end{align}

It is worth noting that for rational approximations obtained by the same
polynomial, in most cases, the closer the orders of the numerator and
denominator polynomials are, the better the approximation effect
\cite{tian2021pade}.

\section{Construction of auxiliary Hamiltonian}\label{auxiliary}

Converting an eigenproblem into a perturbative problem needs an auxiliary
potential $U$, as in Eq. (\ref{2.19}). The auxiliary potential $U$ can be
chosen arbitrarily since $H=\mathcal{H}(1)$ regardless of how $U$ is chosen. A
good choice of auxiliary potential should improve computational efficiency.
This section will provide systematic procedures for constructing auxiliary potentials.

\subsection{Auxiliary potential: Taylor expansion}\label{Taylor}

Consider potentials that can be Taylor expanded.

(1) Select an energy $E$ within the range of interest. The choice of $E$
determines which energy levels will have higher precision in the results. If
interested in a higher excited state, select a larger value of $E$; otherwise,
select a smaller one. For example, if choosing $E=0$, the calculated
eigenvalue with high precision will be near the ground state.

\begin{figure}[ptb]
\centering
%Requires \usepackage{graphicx}
\includegraphics[width=0.48\textwidth]{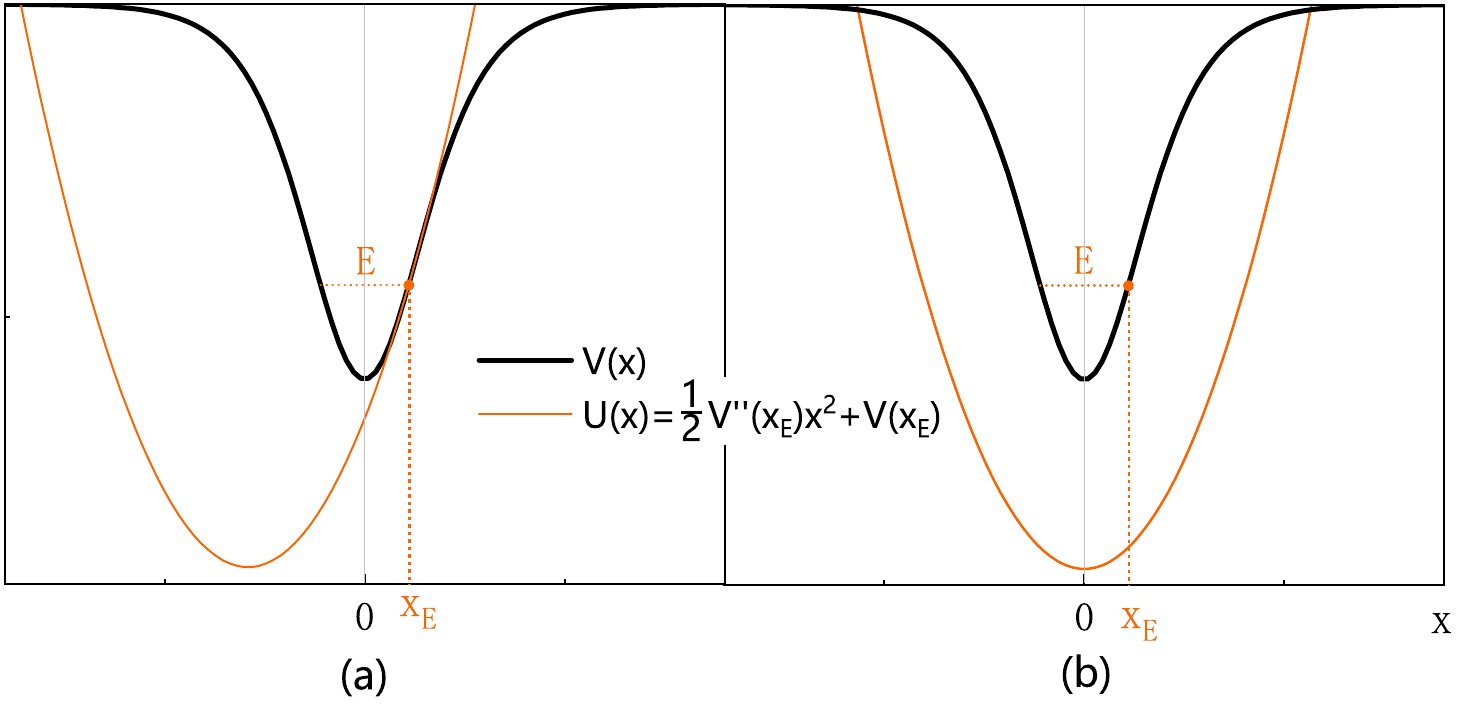}\newline%
\caption{Construction of the auxiliary potential: (a) Expanding the potential
$V\left(  x\right)  $ around $x=x_{E}$; (b) Obtaining the auxiliary potential
$U\left(  x\right)  =\frac{1}{2}V^{\prime\prime}\left(  x_{E}\right)
x^{2}+V\left(  x_{E}\right)  $ by a shifting.}%
\label{sk}%
\end{figure}

(2) Expanding the potential $V\left(  x\right)  $ around the chosen energy
$E$, up to the $2$nd-order term,%
\begin{align}
&  V\left(  x\right)  \simeq V\left(  x_{E}\right)  +V^{\prime}\left(
x_{E}\right)  x+\frac{1}{2}V^{\prime\prime}\left(  x_{E}\right)
x^{2}\nonumber\\
&  =\frac{1}{2}V^{\prime\prime}\left(  x_{E}\right)  \left(  x+\frac
{V^{\prime}\left(  x_{E}\right)  }{V^{\prime\prime}\left(  x_{E}\right)
}\right)  ^{2}+V\left(  x_{E}\right)  -\frac{V^{\prime}\left(  x_{E}\right)
^{2}}{2V^{\prime\prime}\left(  x_{E}\right)  }, \label{V2}%
\end{align}
where $x_{E}$ is given by $V\left(  x_{E}\right)  =E$, as shown in Figure
\ref{sk}a, gives a harmonic oscillator potential with the vertex at $\left(
-\frac{V^{\prime}\left(  x_{E}\right)  }{V^{\prime\prime}\left(  x_{E}\right)
},V\left(  x_{E}\right)  -\frac{V^{\prime}\left(  x_{E}\right)  ^{2}%
}{2V^{\prime\prime}\left(  x_{E}\right)  }\right)  $.

(3) Shifting the vertex of the harmonic oscillator potential (\ref{V2}) along
the $x$-axis to $x=0$ results in a harmonic oscillator potential with vertex
at $\left(  0,V\left(  x_{E}\right)  -\frac{V^{\prime}\left(  x_{E}\right)
^{2}}{2V^{\prime\prime}\left(  x_{E}\right)  }\right)  $: $\frac{1}%
{2}V^{\prime\prime}\left(  x_{E}\right)  x^{2}+V\left(  x_{E}\right)
-\frac{V^{\prime}\left(  x_{E}\right)  ^{2}}{2V^{\prime\prime}\left(
x_{E}\right)  }$. (Note that shifting the harmonic oscillator potential
(\ref{V2}) does not affect the energy level, so it can be performed or omitted
at discretion if only care about the eigenvalue.) The constant $V\left(
x_{E}\right)  -\frac{V^{\prime}\left(  x_{E}\right)  ^{2}}{2V^{\prime\prime
}\left(  x_{E}\right)  }$ in the potential can be arbitrarily chosen. Taking
the zero point of the potential to be $V\left(  x_{E}\right)  $ gives the
auxiliary potential:%
\begin{equation}
U\left(  x\right)  =\frac{1}{2}V^{\prime\prime}\left(  x_{E}\right)
x^{2}+V\left(  x_{E}\right)  , \label{2.6}%
\end{equation}
shown in Figure \ref{sk}b. The auxiliary potential constructed in this way is
a harmonic oscillator potential, which has exact solutions and facilitates
perturbation theory. 

According to Eq. (\ref{2.19}), we rewrite the Hamiltonian (\ref{H}) into a
perturbative form:%
\begin{align}
\mathcal{H}\left(  \lambda\right)   &  =\left[  -\frac{\hbar^{2}}{2\mu}%
\nabla^{2}+\frac{1}{2}V^{\prime\prime}\left(  x_{E}\right)  x^{2}+V\left(
x_{E}\right)  \right] \nonumber\\
&  +\lambda\left[  V-\frac{1}{2}V^{\prime\prime}\left(  x_{E}\right)
x^{2}-V\left(  x_{E}\right)  \right]  . \label{HLamdat}%
\end{align}

\textit{P\"{o}schl-Teller potential.} As an example, we illustrate our method
using the exactly solvable P\"{o}schl-Teller potential. The P\"{o}schl-Teller
potential is ($\frac{\hbar^{2}}{2\mu}=1$) \cite{fl1999practical}%
\begin{equation}
V(x)=-\beta\left(  \beta+1\right)  \operatorname{sech}^{2}x, \label{3.2.1}%
\end{equation}
whose eigenvalue is $E_{n}=-\frac{\hbar^{2}}{2\mu}\left(  \beta-n\right)
^{2}$ with $n=0,1,\ldots,\beta$.

Choose $E$ in the region of interest and construct the auxiliary Hamiltonian
according to Eq. (\ref{HLamdat}):%
\begin{equation}
\mathcal{H}^{\text{PT}}\left(  \lambda\right)  =H_{0}+\lambda\Delta.
\end{equation}
Here $H_{0}=-\nabla^{2}+\beta(\beta+1)\operatorname{sech}^{2}x_{E}\left[
\left(  2\tanh^{2}x_{E}-1\right)  x^{2}-1\right]  $ is an exactly solvable
harmonic oscillator potential with the eigenvalue $E_{n}=\varepsilon
^{2}\left(  2n+1\right)  $ and the eigenfunction $\psi_{n}\left(  x\right)
=\left(  \frac{\varepsilon}{2^{n}n!\sqrt{\pi}}\right)  ^{1/2}\operatorname{H}%
_{n}(\varepsilon x)\exp\left(  -\frac{1}{2}\varepsilon^{2}x^{2}\right)  $,
where $\varepsilon=\left[  \frac{1}{2}V^{\prime\prime}\left(  x_{E}\right)
\right]  ^{1/4}$ and $\operatorname{H}_{n}(x)$ is the Hermitian polynomial.
The perturbative part $\Delta=-\beta\left(  \beta+1\right)
\operatorname{sech}^{2}x_{E}\left[  \frac{\operatorname{sech}^{2}%
x}{\operatorname{sech}^{2}x_{E}}+\left(  2\tanh^{2}x_{E}-1\right)
x^{2}+1\right]  $.

\begin{figure}[ptb]
\centering
%Requires \usepackage{graphicx}
\includegraphics[width=0.48\textwidth]{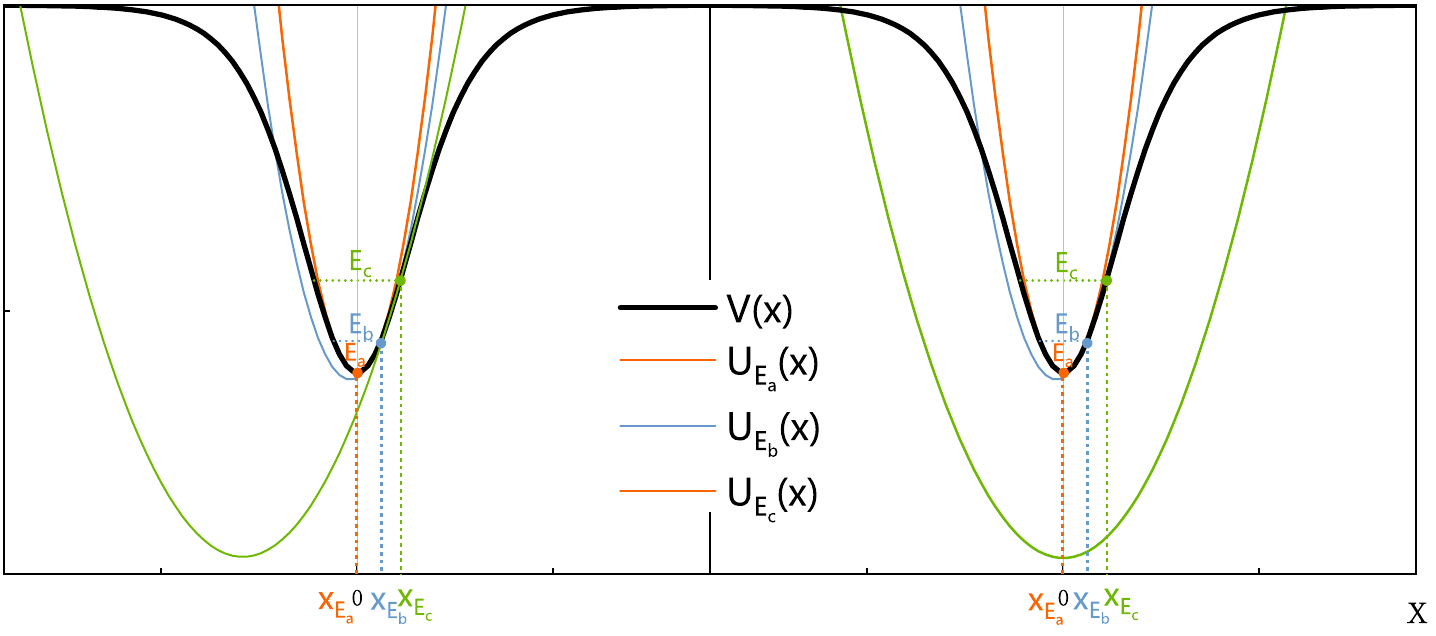}\newline%
\caption{Construction of auxiliary potentials for the P\"{o}schl-Teller
potential ($E_{a}<E_{b}<E_{c}$).}%
\label{Vabc}%
\end{figure}

We choose three values of energy $E$ and obtain the positions of the three
expansion points, $x_{E_{a}}=0$, $x_{E_{b}}=0.3$, $x_{E_{c}}=0.55$, according
to $V\left(  x_{E_{i}}\right)  =E_{i}$. The auxiliary potentials are obtained
by expanding and shifting at these three points $x_{E_{a}}$, $x_{E_{b}}$, and
$x_{E_{c}}$, as shown in Figure \ref{Vabc}.

\begin{figure}[ptb]
\centering
%Requires \usepackage{graphicx}
\includegraphics[width=0.48\textwidth]{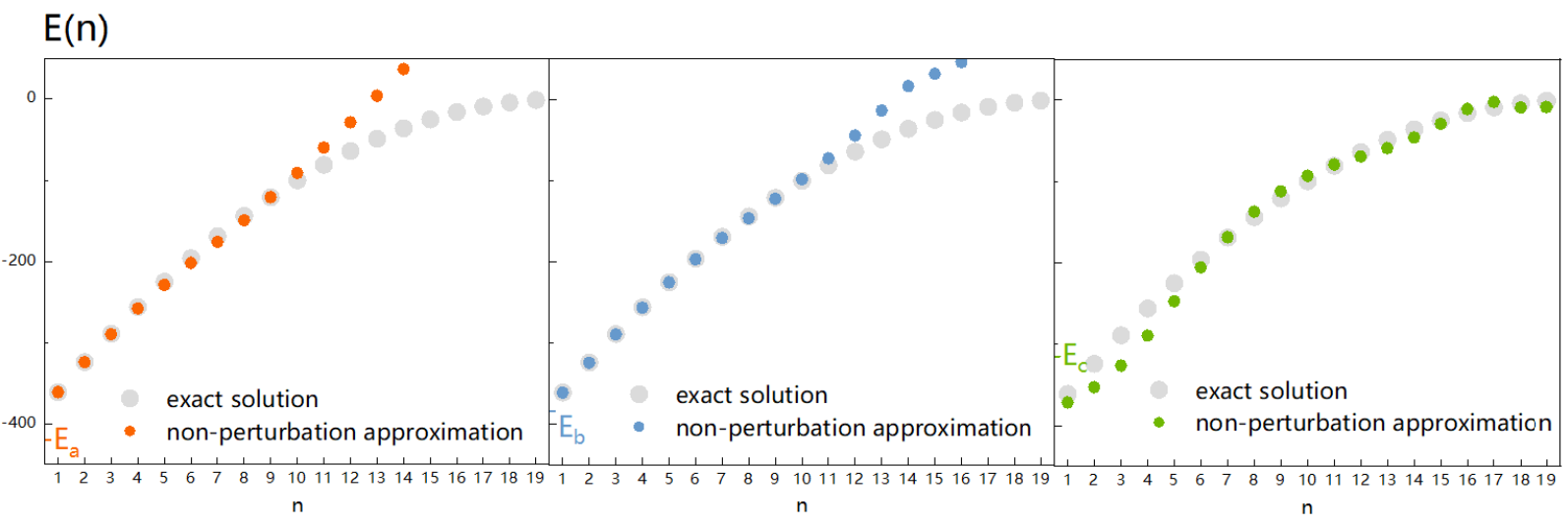}\newline\caption{Comparison of
the non-perturbative eigenvalues of the P\"{o}schl-Teller potential given by
different auxiliary potentials ($E_{a}<E_{b}<E_{c}$) with the exact solution.}%
\label{Eabc}%
\end{figure}

Eq. (\ref{EnPade}) is an explicit expression for the eigenvalue. In Figure
\ref{Eabc}, we plot $E_{n}\simeq\mathcal{E}_{n}^{\left[  8/8\right]  }\left(
1\right)  =\frac{P_{\mathcal{E}}^{[8/8]}(1)}{Q_{\mathcal{E}}^{[8/8]}(1)}$ with
$\beta=20$, for $E_{a}<E_{b}<E_{c}$.

It can be seen from Figure \ref{Eabc} that the eigenvalues corresponding to
$E=0$ are highly accurate near the ground state, while those corresponding to
larger $E$ are highly accurate in higher excited states. To obtain more
eigenvalues, we can construct more auxiliary potentials corresponding to
different $E$.

\begin{figure}[ptb]
\centering
%Requires \usepackage{graphicx}
\includegraphics[width=0.48\textwidth]{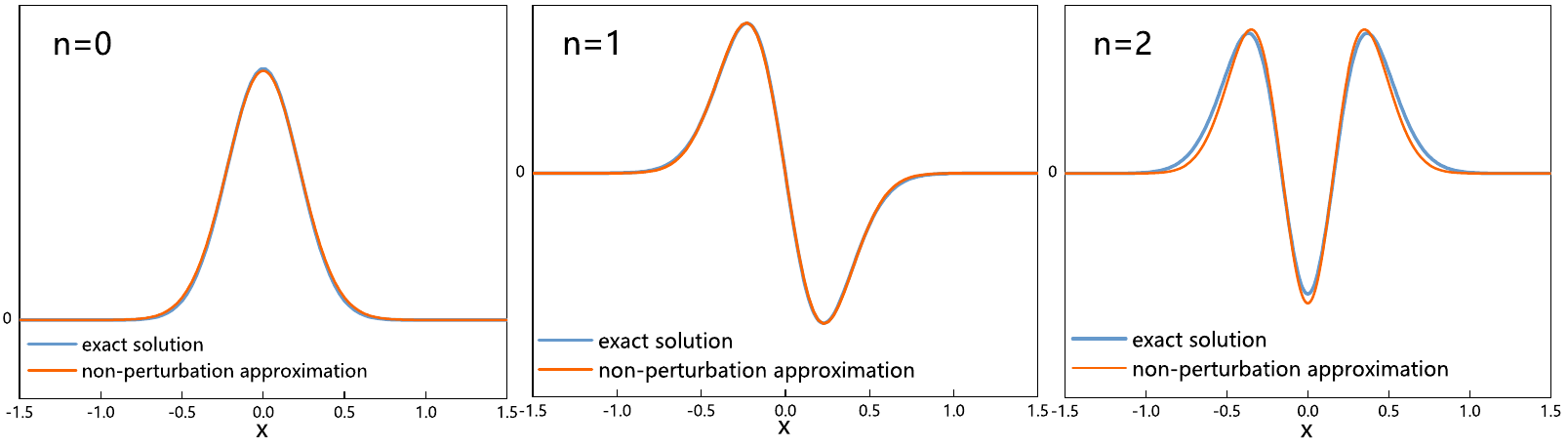}\newline\caption{Comparison of
the non-perturbative eigenfunctions of the P\"{o}schl-Teller potential,
$\psi_{0}$, $\psi_{1}$, and $\psi_{2}$, given by the auxiliary potential
corresponding to $x_{E}=0$ with the exact solution.}%
\label{psi12}%
\end{figure}

Similarly, Eq. (\ref{psiPade}) is an explicit expression for the
eigenfunction. In Figure \ref{psi12}, we plot $\psi_{n}\simeq\psi_{n}%
^{[8/8]}\left(  1\right)  =\frac{P_{\psi}^{[8/8]}(1)}{Q_{\psi}^{[8/8]}(1)}$
with $\beta=20$, for $n=0$, $1$, $2$.

\subsection{Auxiliary potential: Laurent expansion}\label{Coml}

Similar treatment also applies to potentials that can be Laurent expanded. We
take the central potential as an example.

Considering that the negative power law potentials $-\frac{\alpha}{r^{n}}$
have bound states only when $n<2$, here we only consider the case where the
leading order of the Laurent expansion is the negative first power, i.e., the
Coulomb potential. In this case, Laurent expanding a central potential
$V\left(  r\right)  $ around $r=0$ gives
\begin{equation}
V\left(  r\right)  =-\frac{\alpha}{r}+\cdots.
\end{equation}
We take the exactly solvable Coulomb potential as the auxiliary potential,%
\begin{equation}
U\left(  r\right)  =-\frac{\alpha}{r}.
\end{equation}
According to Eq. (\ref{2.19}), the auxiliary Hamiltonian reads%
\begin{equation}
\mathcal{H}\left(  \lambda\right)  =\left[  -\frac{\hbar^{2}}{2\mu}\nabla
^{2}-\frac{\alpha}{r}\right]  +\lambda\left[  V\left(  r\right)  +\frac
{\alpha}{r}\right]  , \label{HLamdar}%
\end{equation}
where $H_{0}=-\frac{\hbar^{2}}{2\mu}\nabla^{2}-\frac{\alpha}{r}$ is the
Coulomb potential.

\textit{Hulthen potential.} As an example, we consider the exactly solvable
spherically symmetric Hulthen potential \cite{fl1999practical},%
\begin{equation}
V\left(  r\right)  =-V_{0}\frac{e^{-r/r_{0}}}{1-e^{-r/r_{0}}}, \label{3.1.1}%
\end{equation}
whose eigenvalue for $l=0$ is $E_{n}=-V_{0}\left(  \frac{\beta^{2}-n^{2}%
}{2\beta n}\right)  ^{2}$ with $\beta^{2}=\frac{2\mu V_{0}}{\hbar^{2}}%
r_{0}^{2}$ and $\beta^{2}-n^{2}>0$.

The leading term of the Laurent expansion of the Hulthen potential
(\ref{3.1.1}) is $-V_{0}\frac{r_{0}}{r}$. Chosing the Coulomb potential,
$U\left(  r\right)  =-V_{0}\frac{r_{0}}{r}$, as the auxiliary potential, the
auxiliary Hamiltonian for $l=0$ by Eq. (\ref{HLamdar}) reads ($\frac{\hbar
^{2}}{2\mu}=1$)%
\begin{align}
\mathcal{H}\left(  \lambda\right)   &  =\left(  -\frac{1}{r}\frac{d^{2}%
}{dr^{2}}r-V_{0}\frac{r_{0}}{r}\right) \nonumber\\
&  +\lambda\left(  -V_{0}\frac{e^{-r/r_{0}}}{1-e^{-r/r_{0}}}+V_{0}\frac{r_{0}%
}{r}\right)  . \label{3.1.5}%
\end{align}
The eigenvalue and eigenfunction of $H_{0}=-\frac{1}{r}\frac{d^{2}}{dr^{2}%
}r-V_{0}\frac{r_{0}}{r}$ are $E_{n}=-\frac{V_{0}^{2}r_{0}^{2}}{4n^{2}}$ and
$\psi_{n}\left(  r\right)  =\frac{\sqrt{2}}{8\pi n^{2}}\left(  r_{0}%
V_{0}\right)  ^{3/2}\sqrt{\frac{n!}{\left(  n-1\right)  !}}e^{-\frac
{r_{0}V_{0}}{2n}r}\left.  {}\right.  _{1}F_{1}\left(  -n+1,2,\frac{r_{0}V_{0}%
}{n}r\right)  $ with $\left.  {}\right.  _{1}F_{1}\left(  a,b,z\right)  $ the
confluent hypergeometric function.

\begin{figure}[ptb]
\centering
%Requires \usepackage{graphicx}
\includegraphics[width=0.45\textwidth]{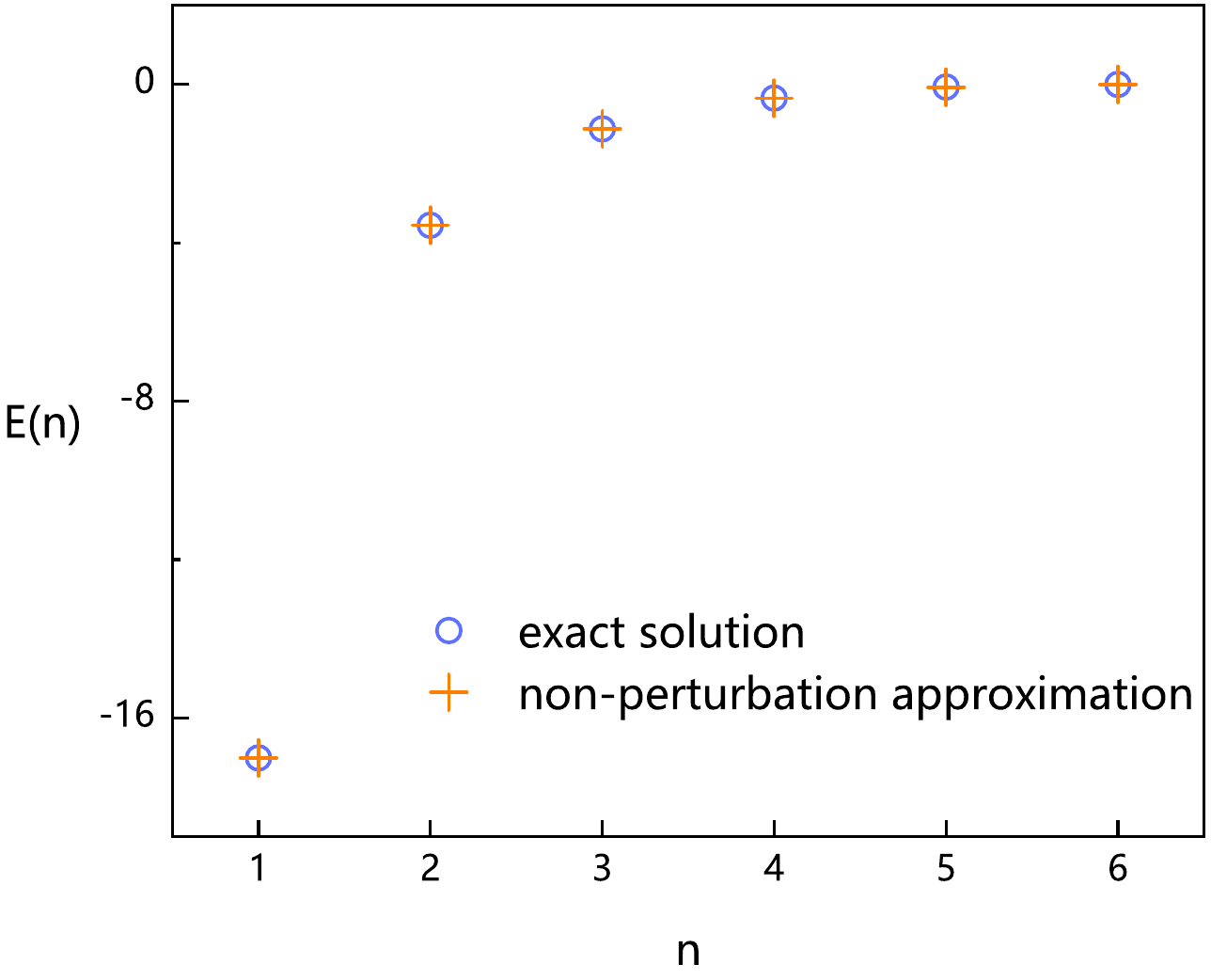}\newline\caption{Comparison of
the non-perturbative result of eigenvalues of the Hulthen potential with the
exact solution.}%
\label{Hulthen}%
\end{figure}

In Fig. \ref{Hulthen}, we plot the eigenvalue given by Eq. (\ref{EnPade}),
$E_{n}\simeq\mathcal{E}_{n}^{\left[  8/8\right]  }\left(  1\right)
=\frac{P^{[8/8]}(1)}{Q^{[8/8]}(1)}$ with $\frac{\hbar^{2}}{2\mu}=1$, $V_{0}%
=2$, and $r_{0}=3$.

\subsection{Auxiliary potential: numerical fitting}\label{Fit}

For the potentials that cannot be Taylor or Laurent expanded, we suggest a
fitting-based scheme for constructing auxiliary potentials.

Based on the form of the potential $V$, we first choose an exactly solvable
auxiliary potential $U$ which should contain some undetermined parameters
$k_{1},k_{2},\cdots$. Then we fit the auxiliary potential $U$ to $V$ to
determine these parameters. After determining the parameters in the auxiliary
potential $U$ through fitting, we can calculate the energy levels using the
method described above. Here, we consider two examples: one uses a linear
potential as the auxiliary potential, and the other uses~a quadratic potential
as the auxiliary potential.

$r^{2/3}$\textit{-potential. }Consider\textit{ }the exactly solvable $r^{2/3}%
$-potential%
\begin{equation}
V\left(  r\right)  =r^{2/3}. \label{3.4.1}%
\end{equation}
For convenience, we only consider the asymptotic solution: $E_{n}\simeq
\frac{4\sqrt{6}}{3}\left(  n-\frac{1}{8}\right)  ^{1/2}$ \cite{chen2022exact}.

Here, we take the exactly solvable linear potential as the auxiliary
potential:%
\begin{equation}
U\left(  r\right)  =kr+r_{0}, \label{3.4.4}%
\end{equation}
where $k\ $and $r_{0}$ are fitting parameters to be determined. The auxiliary
Hamiltonian reads ($\frac{\hbar^{2}}{2\mu}=1$ and $l=0$):%
\begin{equation}
\mathcal{H}\left(  \lambda\right)  =\left(  -\frac{1}{r}\frac{d^{2}}{dr^{2}%
}r+kr+r_{0}\right)  +\lambda\left(  r^{2/3}-kr-r_{0}\right)  . \label{3.4.6}%
\end{equation}
The asymptotic solution of eigenvalue of $H_{0}=-\frac{1}{r}\frac{d^{2}%
}{dr^{2}}r+kr+r_{0}$ is $E_{n}=\left(  3\pi k/2\right)  ^{2/3}\left[  \left(
2n-1/2\right)  \right]  ^{2/3}+r_{0}$, the eigenfunction is $\psi_{n}\left(
r\right)  =C_{n}\operatorname{Ai}\left(  \xi\right)  $ with $\operatorname{Ai}%
\left(  \xi\right)  $ the Airy function, $\xi=2^{1/3}k^{-2/3}\left(
kr-E_{n}\right)  $, $C_{n}=\left(  2k\right)  ^{1/6}/\sqrt{\alpha
_{n}\operatorname{Ai}\left(  -\alpha_{n}\right)  ^{2}+\operatorname{Ai}%
^{\prime}\left(  -\alpha_{n}\right)  ^{2}}$, and $\alpha_{n}=2^{1/3}%
r_{0}k^{-2/3}+\frac{1}{2}\left[  3\pi\left(  4n-1\right)  \right]  ^{2/3}$
\cite{fl1999practical}.

\begin{figure}[ptb]
\centering
%Requires \usepackage{graphicx}
\includegraphics[width=0.48\textwidth]{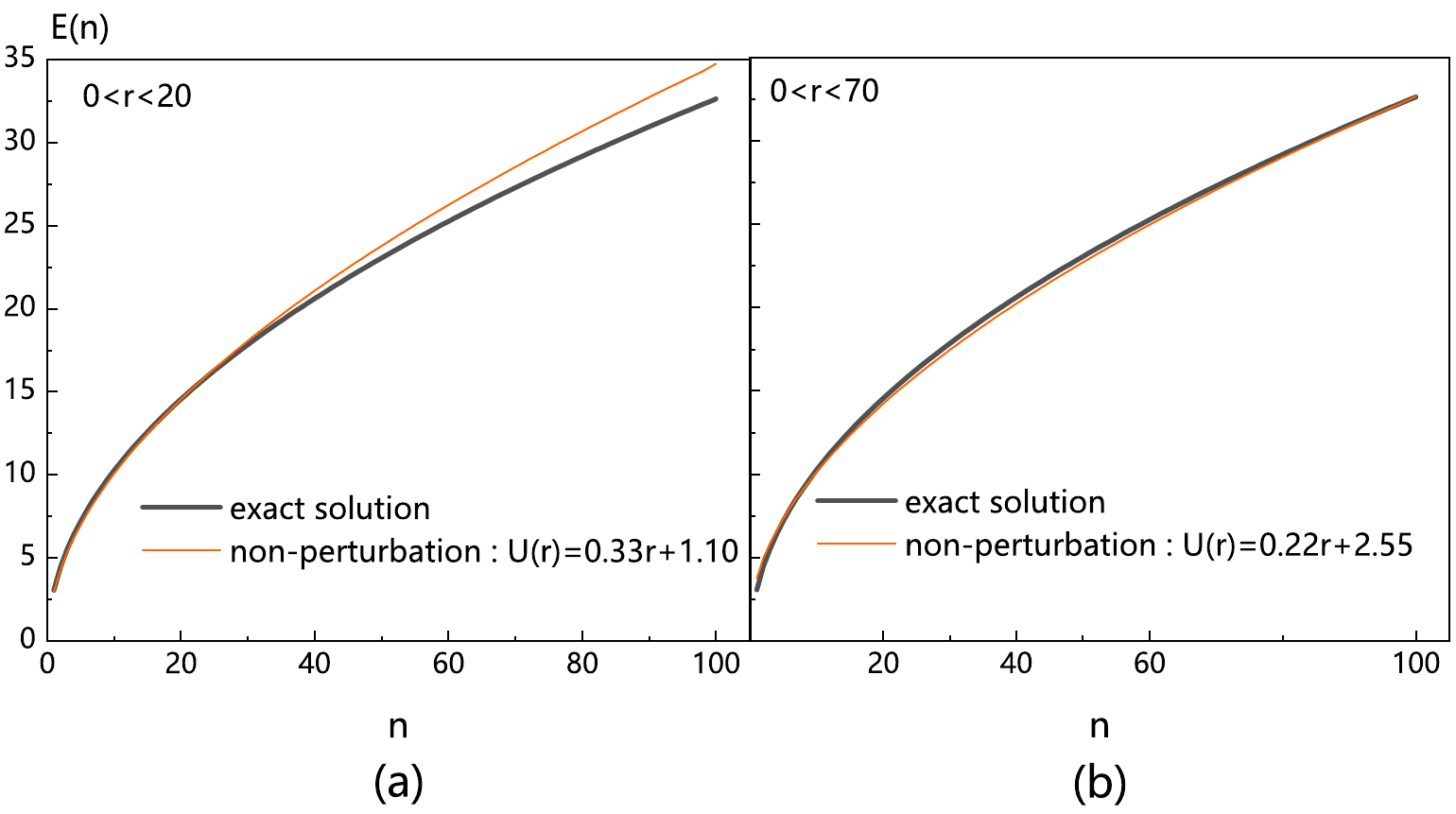}\newline%
\caption{Comparison of the non-perturbative result of eigenvalues of the
$\frac{2}{3}$-potential with the asymptotic solution.}%
\label{23split}%
\end{figure}

Different fitting ranges give different auxiliary potentials with different
values of $k$ and $r_{0}$. In Figure \ref{23split}, we plot $E_{n}%
\simeq\mathcal{E}_{n}^{\left[  8/8\right]  }\left(  1\right)  =\frac
{P^{[8/8]}(1)}{Q^{[8/8]}(1)}$ corresponding to two auxiliary potentials: (a)
fitting range $0<r<20$ gives the auxiliary potential $U\left(  r\right)
=0.331619r+1.10428$; (b) fitting range $0<r<70$ gives the auxiliary potential
$U\left(  r\right)  =U\left(  r\right)  =0.2184r+2.54714$. It can be seen that as the fitting range expands, the region of
high precision shifts from low to high excited states.

\textit{Flat-bottom potential.} For the flat-bottom potential,%
\begin{equation}
V\left(  x\right)  =\left\{
\begin{array}
[c]{c}%
\left(  x-L\right)  ^{2}\text{ \ \ \ \ }x>L\\
0\text{ \ \ \ \ \ \ \ \ \ \ }\left\vert x\right\vert <L\\
\text{ }\left(  x+L\right)  ^{2}\text{ \ \ \ \ }x<-L
\end{array}
\right.  , \label{fbP}%
\end{equation}
we use an exactly solvable quadratic potential as the auxiliary potential:
\begin{equation}
U\left(  x\right)  =x^{2},
\end{equation}
whose eigenvalue $\mathcal{E}_{n}^{\left(  0\right)  }=2\left(  n+\frac{1}%
{2}\right)  $ and eigenfunction $\psi_{n}^{\left(  0\right)  }=\left(
\frac{1}{2^{n}n!\sqrt{\pi}}\right)  ^{1/2}\operatorname{H}_{n}\left(
x\right)  \exp\left(  -\frac{1}{2}x^{2}\right)  $.

\begin{figure}[ptb]
\centering
%Requires \usepackage{graphicx}
\includegraphics[width=0.52\textwidth]{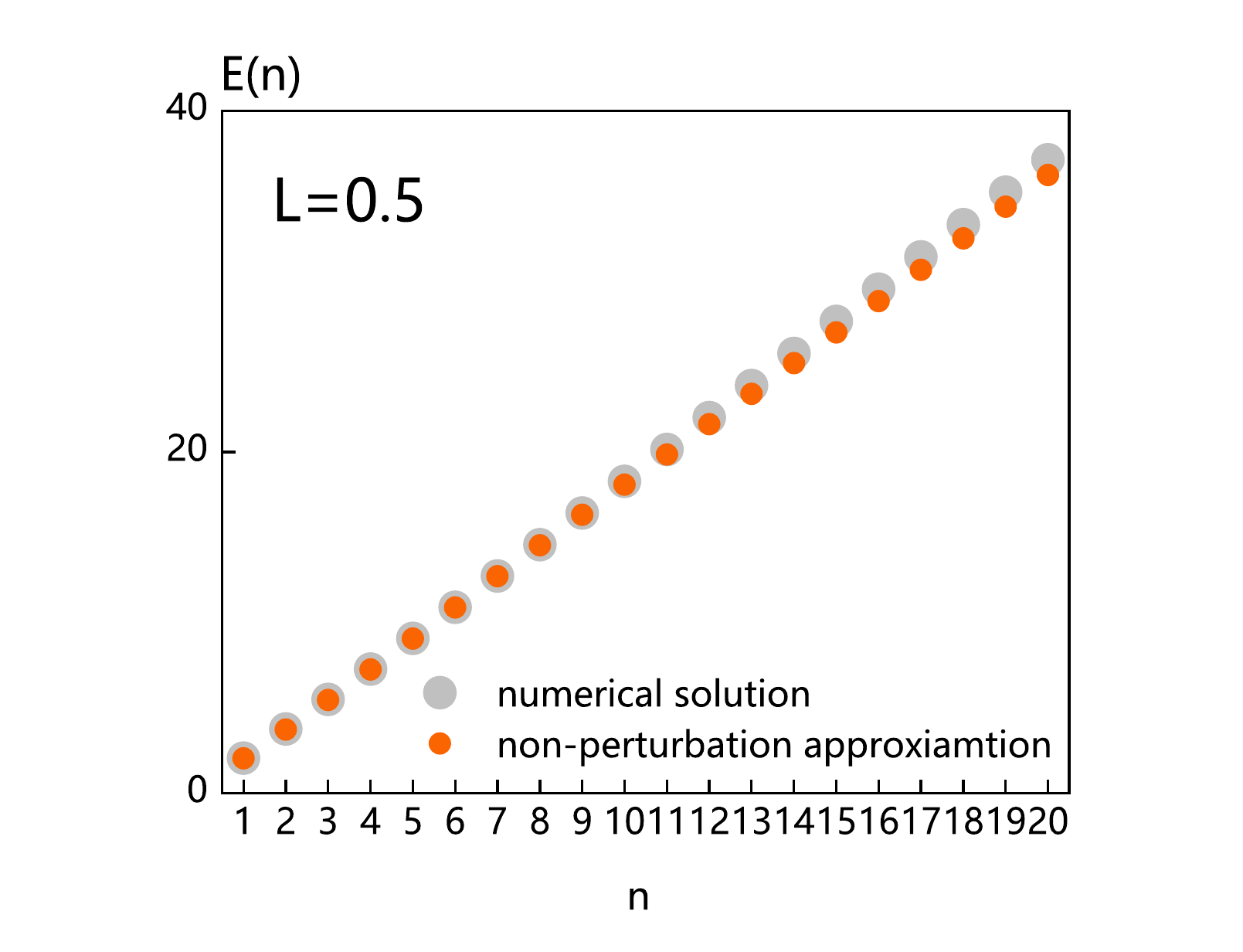}\newline\caption{Comparison of the
non-perturbative result of eigenvalues of the flat-bottom potential with the
numerical solution.}%
\label{flatfig}%
\end{figure}

The flat-bottom potential (\ref{fbP}) lacks an exact solution; we compare our
non-perturbative result $E_{n}\simeq\mathcal{E}_{n}^{\left[  8/8\right]
}\left(  1\right)  =\frac{P^{[8/8]}(1)}{Q^{[8/8]}(1)}$ with numerical
solutions in Figure \ref{flatfig}.\qquad

\section{Conclusion}\label{Conclusion}

In this paper, we propose a nonperturbative method for solving the
eigenproblem of the Hamiltonian $H=T+V$. This method, like perturbation
theory, has a standardized procedure. The nonperturbative explicit expressions
of the eigenvalue and eigenfunction are given.

The basic idea of the method is to first (1) convert a nonperturbative problem
into a perturbative one and (2) then analytically continue the perturbative
result to the nonperturbative region. We use, but in principle not limited to,
the rational approximation to perform approximative analytic continuations. As
shown in the examples, this method yields very good results for a general
class of potentials.

Compared with another nonperturbative method, the variational method, we can
see that (1) the variational method is based on guessing trial wave functions,
which cannot be automatically carried out according to a standardized
procedure, while our method has a standardized procedure where the calculation
proceeds automatically, like in perturbation theory, and (2) the variational
method is only effective for the ground and low-excited states, while our
method is effective for arbitrary energy levels.

We demonstrate this nonperturbative method using quantum mechanical
eigenproblems. The idea of this method can be applied to any field where
perturbative treatments are well-established, such as quantum field theory and
statistical mechanics. In our future work, we will use this method to convert
perturbative techniques in quantum field theory, such as thermal field theory
and Feynman diagrammatic expansions, into nonperturbative methods.

Perturbation theory is ineffective for strong coupling problems (coupling
constants greater than $1$). Perturbation theory results in a perturbation
series, or more strictly speaking, a truncated series with a finite number of
terms. Perturbation theory gives a power series of the coupling constant, and
the effective range of the perturbation series is within the convergence
circle of the power series. The necessary condition for convergence of the
power series is that the coupling constant must be smaller than $1$. Thus, (1)
perturbation theory is not applicable for coupling constants greater than $1$;
(2) even if the coupling constant is smaller than $1$, the perturbation series
may not converge (because the condition that the coupling constant must be
smaller than $1$ is necessary but not sufficient). In contract, the rational
function approximation obtained by analytic continuing perturbation theory not
only applies at $\lambda=1$ but is also valid when $\lambda>1$. We only use
$\lambda=1$ in the problem under consideration. This method can also be used
to deal with nonperturbative strong coupling problems with $\lambda>1$: first
treating a strong coupling problem as a weak coupling problem which can be
solved by perturbation theory, and then obtaining the strong coupling result
through analytic continuation. In the strong interaction, because the coupling
constant's magnitude is unsuitable for perturbation theory, the calculation of
the strong interaction is greatly limited. The nonperturbative method
presented in this paper can be directly applied to the strong interaction
theory whose coupling constant is near or larger than $1$.

Power series approximation often appears as a polynomial approximation, which
loses information about singularities for the polynomial has no singularities.
The rational approximation can partially recover information about
singularities: the poles of the rational function are approximations of
singularities in the original function.

%----------------------------------------------------------

\section*{Acknowledgments}
\textsf{\small{We are very indebted to Dr G. Zeitrauman for his encouragement. This work is supported in part
by Special Funds for theoretical physics Research Program of the NSFC under Grant No. 11947124,
and NSFC under Grant Nos. 11575125 and 11675119.}}

\bigskip

%\printbibliography[title=References]
%

%\bibliographystyle{JHEP}
%\bibliography{refs.bib}

%----------------------------------------------------------

  % Divirta-se escrevendo com a classe  \LaTeX\  \cls{rctart} \hspace{5pt}\faHandSpock[regular] \\%\faChessKnight \\ \faLinux
%     \noindent\faWix\hspace{5pt}\href{https://memonotess1.wixsite.com/memonotess}{https://memonotess1.wixsite.com/memonotess} \\

\bigskip
\bigskip
    \noindent\faEnvelope[regular]\hspace{7pt}\href{mailto://silvio.granja@unemat.br}{liwendu@tjnu.edu.cn} (W-D Li)\\
    \hspace{7pt}\faEnvelope[regular]\hspace{7pt}\href{mailto://silvio.granja@unemat.br}{daiwusheng@tju.edu.cn} (W-S Dai)\\
   % \faUniversity\hspace{7pt}\href{http://unemat.br}{http://unemat.br}\\
    %\faMapMarked\hspace{7pt}\href{https://maps.app.goo.gl/d1oovfssBBWaKnRk6}{http://unemat.br}
%     \faEnvelope[normal]\hspace{7pt}eduardo.gracidas29@gmail.com \\
%     \faInstagram\hspace{8pt}memo.notess

\end{document}